\documentclass[aps,epsfig,graphics,floatfix,mathbbm,twocolumn,a4paper]{revtex4}

\usepackage{amsmath,amsfonts,amssymb,graphics,graphicx,epsfig,color,times}

\begin{document}
\bibliographystyle{apsrev}

\newcommand{\R}{\mathbbm{R}}
\newcommand{\rr}{\mathbbm{R}}
\newcommand{\nn}{\mathbbm{N}}
\newcommand{\cc}{\mathbbm{C}}
\newcommand{\ii}{\mathbbm{1}}
\newcommand{\M}{{\cal M}}
\newcommand{\T}{{\cal T}}
\newcommand{\1}{\mathbbm{1}}
\newcommand{\id}{{\rm id}}
\newcommand{\C}{{\cal C}}
\newcommand{\Tt}{{\cal T}}
\newcommand{\U}{{\cal U}}
\newcommand{\tr}{{\rm tr}}
\newcommand{\gr}[1]{\boldsymbol{#1}}
\newcommand{\be}{\begin{equation}}
\newcommand{\ee}{\end{equation}}
\newcommand{\bea}{\begin{eqnarray}}
\newcommand{\eea}{\end{eqnarray}}
\newcommand{\ket}[1]{|#1\rangle}
\newcommand{\bra}[1]{\langle#1|}
\newcommand{\avr}[1]{\langle#1\rangle}
\newcommand{\G}{{\cal G}}
\newcommand{\eq}[1]{Eq.~(\ref{#1})}
\newcommand{\ineq}[1]{Ineq.~(\ref{#1})}
\newcommand{\sirsection}[1]{\section{\large \sf \textbf{#1}}}
\newcommand{\sirsubsection}[1]{\subsection{\normalsize \sf \textbf{#1}}}
\newcommand{\ack}{\subsection*{\normalsize \sf \textbf{Acknowledgements}}}
\newcommand{\front}[5]{\title{\sf \textbf{\Large #1}}
\author{#2 \vspace*{.4cm}\\
\footnotesize #3}
\date{\footnotesize \sf \begin{quote}
\hspace*{.2cm}#4 \end{quote}
#5} \maketitle}
\newcommand{\eg}{\emph{e.g.}~}

\newcommand{\proofend}{\hfill\fbox\\\medskip }


\newtheorem{theorem}{Theorem}
\newtheorem{proposition}{Proposition}

\newtheorem{lemma}{Proposition}

\newtheorem{definition}{Definition}
\newtheorem{corollary}{Corollary}
\newtheorem{example}{Example}

\newcommand{\proof}[1]{{\bf Proof.} #1 $\proofend$}

\title{Simplifying additivity problems using direct sum constructions}

\author{Motohisa Fukuda$^1$, Michael M. Wolf$^2$}
\affiliation{$^1$Statistical Laboratory, Centre for Mathematical
Sciences, University of Cambridge\\
$^2$ Max-Planck-Institute for Quantum Optics,
 Hans-Kopfermann-Str.\ 1, D-85748 Garching, Germany.}

\date{\today}


\begin{abstract}
We study the additivity problems for the classical
capacity of quantum channels, the minimal output entropy and its
convex closure. We show for each of them that additivity for
arbitrary pairs of channels holds iff it holds for arbitrary equal
pairs, which in turn can be taken to be unital. In a similar
sense, weak additivity is shown to imply strong additivity for any
convex entanglement monotone. The implications are obtained by
considering direct sums of channels (or states) for which we show
how to obtain several information theoretic quantities from their
values on the summands. This provides a simple and general tool
for lifting additivity results.
\end{abstract}

\maketitle



\section{Introduction}

A central question in classical and quantum information theory is,
how much information can be transmitted through a given noisy
channel. For classical channels the maximal asymptotically
achievable rate---the capacity---was derived in the seminal work
of Shannon \cite{Shannon}. For quantum channels, however, the
matter is complicated by the existence of entanglement and the
possibility of exploiting it in the encoding to protect
information against decoherence. If one excludes this possibility,
a capacity formula for the transmission of classical information
through quantum channels was proven by Holevo \cite{Holevo} and
Schumacher and Westmoreland \cite{SW} (HSW). Since then,
considerable effort was devoted to the question whether (or in
which cases) entangled inputs can lead to rates beyond the HSW
capacity. This issue---the \emph{additivity problem} for the HSW
capacity---is still undecided, although for several classes of
channels additivity has been shown to be true, i.e., entanglement
does not seem to help in any case (see, e.g.,
\cite{WE,King1,King2,Shor} and references therein). Instead, other
additivity problems appeared which are similar in spirit but
concern very different quantities like the minimal output entropy
and the entanglement of formation, an entanglement measure for
bipartite states for which in addition strong super-additivity has
been conjectured.

A major conceptional insight was then gained in
\cite{Sho03,Pom03,AB04,MATWIN} where it was shown that all these
additivity problems are globally equivalent in the sense that if
additivity holds for one of these quantities in general, then it
does so for all of them. Here `in general' means that it has to be
true for arbitrary pairs of channels (or states), a condition we
will call \emph{strong additivity}.

In this work we present a further conceptional simplification of
these and related additivity problems. We show that strong
additivity is implied by \emph{weak additivity}, meaning
additivity for arbitrary pairs of \emph{equal} channels or states.
Moreover, based on \cite{Fuk06} we argue that it suffices to
consider pairs of identical unital channels only. 
This observation may be a small
step on a notorious path but it might guide future research as it
for instance underlines recent attempts to understand the
asymptotic structure of tensor powers of unital channels
\cite{Birkhoff}. Moreover, one may think of other additivity
questions than the ones stated above for which our techniques
could be of use. In particular, we think of regularized quantities
(like quantum capacities (cf. \cite{WP,SMWI}) or certain
entanglement measures) for which weak additivity holds by
definition.
 
Our main tool is the use of direct sums of channels or states. For
the latter case similar constructions appeared in
\cite{tag1,tag2,Plenio}. We begin with a discussion of direct sum
channels. This will contain more than what is needed for the
subsequent additivity results as we think that these tools might
be of independent interest.

\section{Direct sums of quantum channels}

We consider direct sums of channels, i.e., completely positive and
trace preserving maps of the form $T=\oplus_i T_i$, where each
$T_i$ is a channel in its own right. Our aim in this section is to
express information theoretic functionals of $T$ in terms of their
values for the $T_i$'s. The definition of the quantities appearing
in the following proposition will be given in the proof.

\begin{lemma}[Direct sums]\label{prop:ds}
Consider a direct sum $T=\oplus_{i=1}^n T_i$, $n\in\mathbb{N}$ of
arbitrary finite dimensional channels. Then
\begin{enumerate}
    \item Minimal output $\alpha$-Renyi entropy ($\alpha\geq
1$):\be\label{eq:Sa0}
    S_{\min,\alpha}\big(T \big)=\min_i\big[S_{\min,\alpha}(T_i)\big],\ \ee
    \item Coherent information:
    \be
    J\big(T\big)=\max_i\big[J(T_i)\big],\label{eq:J0}\ee
    \item Mutual information:
    \bea I\big(T\big)&=&\max_{\{\lambda_i\}}\
    S(\{\lambda_i\}) + \sum_i\lambda_i I(T_i)\label{eq:I0},\\ \label{eq:I1}
    &=& \log \sum_i 2^{I(T_i)},\eea
    where $\{\lambda_i\}$ is a probability distribution and
    $S(\{\lambda_i\})$ its entropy.
    \item HSW capacity:
    \bea\chi\big(T\big)&=& \max_{\{\lambda_i\}}\
    S(\{\lambda_i\}) + \sum_i\lambda_i
    \chi(T_i)\label{eq:chi0}\\
    &=& \log \sum_i 2^{\chi(T_i)}. \label{eq:chi1}\eea
\end{enumerate}
\end{lemma}\newpage

{\bf Remark:} Let us briefly comment on the interpretation of the
above formulas. Concerning the HSW capacity, classical information
can either be sent through the channels $T_i$ or it can be encoded
in the choice of blocks $i=1,\ldots,n$. Eq.(\ref{eq:chi0}) shows
exactly the competition between these two ways of communicating
classical information. For the quantum mutual information, which
gives the entanglement assisted capacity \cite{Eassisted}, we
obtain the same interpretation (note that the '2' comes from the
fact that we take $\log$ in base 2). The coherent information is
related (via regularization) to the quantum capacity \cite{QCap}.
In this case encoding information in the choice of blocks is not
possible---this would be purely classical as all the coherences
get lost. Similarly, for the minimal output entropies the minimum
is obtained by putting all the weight into the least noisy
channel.

\proof{1.
The \emph{$\alpha$-Renyi entropy} is defined as
\be
 S_\alpha (\rho)=
\frac{1}{1-\alpha} \log {\rm tr} [\rho^\alpha]
=\frac{\alpha}{1-\alpha}\log \|\rho\|_\alpha, \ee for $0
\leq \alpha \leq \infty$. Here $\|\cdot\|_p$ is the Schatten
$p$-norm. When $\alpha = 1$ the functional is defined by its limit
which is the \emph{minimal output entropy} $S_{\min}(T)=\inf_\rho
S(T(\rho))$ with $S(\rho)=-{\rm tr}\rho \log \rho$ the von Neumann
entropy. Let us consider this case first.

 As the direct sum
$\oplus_i T_i$ erases the off-diagonal blocks so that all possible
outputs can be obtained upon block-diagonal inputs we can restrict
to $\rho=\oplus_i\tilde{\rho}_i$. Here $\tilde{\rho}_i$ is not
necessarily normalized so that the weights
$\tr[\tilde{\rho}_i]=:\lambda_i$ form a probability distribution.
Writing $\rho_i:=\tilde{\rho}_i/\lambda_i$ and using the concavity
of von Neumann entropy we get \be S\big(\oplus_i
T_i(\lambda_i\rho_i)\big) \geq \sum_i \lambda_i S(T_i(\rho_i)).
\ee This leads to Eq.(\ref{eq:Sa0}) when $\alpha=1$. For $\alpha\
> 1$ the minimization of $S_\alpha(T(\rho))$ amounts to a maximization of 
$\|T(\rho)\|_\alpha$ and
 the result follows from convexity of $\|\cdot\|_\alpha$ in a
similar way. \vspace*{5pt}

2. The \emph{coherent information} is defined as
\be
J(T)=\sup_\rho S\big(T(\rho)\big)-S\big(T\otimes\id(\Psi)\big),\
\ee
where $\Psi$ is a purification of $\rho$ such that $\rho=\tr_B\Psi$.
Since $T$ and $T \otimes \id$ erase the off-diagonal blocks
we can replace $\rho$ and $\Psi$
by their diagonal blocks:
$\oplus_i \lambda_i \rho_i$ and 
$\oplus_i \lambda_i \Psi_i$.
Here, $\Psi_i$ is an extension of $\rho_i$.
Since the conditional entropy
$S(\rho_{AB})- S(\rho_A)$ is concave in $\rho_{AB}$
\cite{Ruskai} 
considering a convex decomposition of each $\Psi_i$ into pure
states shows Eq.(\ref{eq:J0}) in a similar way as above. 
\vspace{5pt}

3. The \emph{mutual information} defined as \be I(T)=\sup_\rho\
S(\rho)+S\big(T(\rho)\big)-S\big(T\otimes\id(\Psi)\big)\ee is
concave in $\rho$ 
so 
the maximum will be achieved by a block diagonal
$\rho= \oplus_i\lambda_i\rho_i$ for $T=\oplus_i T_i$.
To see this,
let $V=\oplus_j \exp \{2\pi{\rm i}j/n \}I_j$ and
average $V^k \rho V^{\ast k}$ over $k=1,\ldots,n$.
Take a purification $\Psi$ of $\rho=\oplus_i\lambda_i\rho_i$,
and then replace $\Psi$ by its diagonal blocks:
$\oplus_i \lambda_i \Psi_i$ as before.
However, each $\Psi_i$ is a purification of $\rho_i$ in this case.
Indeed, suppose $\rho_i=\sum_j p_{ij}|ij \rangle\langle ij|$,
where $\{ |ij\rangle \}_j$ is an orthonormal basis 
in the $i$th subspace.
Then, $\Psi$ is
\begin{align*}
\sum_{ijkl} \sqrt{ \lambda_i \lambda_k  p_{ij}p_{kl}} 
|ij\rangle\langle kl| \otimes |ij\rangle \langle kl|,
\end{align*}
and its $i$th diagonal block $\lambda_i\Psi_i$ is
\begin{align*}
\lambda_i \sum_{jl}\sqrt{p_{ij}p_{il}} 
|ij\rangle\langle il| \otimes |ij\rangle \langle il|
= \lambda_i |\Psi_i\rangle\langle \Psi_i|.
\end{align*}
Here, $|\Psi_i \rangle = \sum_j  \sqrt{p_{ij}}|ij\rangle \otimes |ij\rangle$.
Exploiting this together with
$S(\lambda \rho)=\lambda (S(\rho)-\log\lambda)$ then gives
Eq.(\ref{eq:I0}). 
Eq.(\ref{eq:I1}) follows then from determining
the optimal $\lambda_i$ via Lagrange multipliers in the following
way. The maximization problem of $S(\{\lambda_i\})+\sum \lambda_i
c_i$ for a probability distribution $\{\lambda_i\}$ amounts then
to maximizing \be S(\{\lambda_i\})+\sum_i \lambda_i c_i + \Lambda
\left(\sum \lambda_i -1\right), \ee where $\Lambda$ is the
Lagrange multiplier. Taking partial derivatives we obtain for
extremal $\{\tilde{\lambda}_i\}$:
\begin{align}
- \log \tilde{\lambda}_i - \frac{1}{\ln 2}
+ c_i + \Lambda &= 0 \qquad \forall i \label{lagrange1}\\
\sum_i \tilde{\lambda}_i - 1 &= 0\label{lagrange2}.
\end{align}
Hence (\ref{lagrange1}) shows $c_i-\log \tilde{\lambda}_i$ is a
constant, say, $C$ for $\forall i$, and by (\ref{lagrange2}) we
get $C=\log\sum_i 2^{c_i}$. Therefore \be S(\{\tilde{\lambda}_i
\}) + \sum_i \tilde{\lambda}_i c_i 
= \sum_i \tilde{\lambda}_i C =
\log \sum_i 2^{c_i}.  \ee As this is lower bounded by $\min_i
[I(T_i)]$ it must be the maximum. \vspace{5pt}

4. The \emph{HSW capacity} $\chi$ is given by 
\begin{align} 
&\chi(T) =
\sup_\rho\
S(T(\rho))- H_{T}(\rho)\label{eq:HSWC},\\
&\text{where}\qquad
H_T(\rho)=\inf_{\sum p_k \rho_k = \rho} \sum_k p_k
S(T(\rho_k)).
\end{align}
Here, $H_T(\rho)$ is the \emph{convex closure of
the output entropy}; $\{p_k \}$ is a probability distribution and
$\rho_k$ are density matrices. The r.h.s. of Eq.(\ref{eq:HSWC})
for a fixed average input state $\rho$ is a \emph{constraint HSW
capacity} which we will denote by $\chi(T,\rho)$. Since the inputs
can again be assumed to be block-diagonal we have:
\begin{align}
\chi\left(\oplus_i T_i, \oplus_i \lambda_i \rho_i\right)
&= S(\oplus_i \lambda_i T_i(\rho_i))
- \sum_i \lambda_i H_{T_i}(\rho_i) \notag\\
&= S(\{\lambda_i\}) + \sum_i \lambda_i \chi(T_i,\rho_i).
\end{align}
The first equality is explained by the fact that since the von
Neumann entropy is concave there is an optimal decomposition of
$\oplus_i \lambda_i \rho_i$ for which each state  has its support
in one of the diagonal blocks. The second equality comes from
$S(\lambda_i \rho_i)= \lambda_i (S(\rho_i)-\log \lambda_i)$.
Taking the supremum over all states $\{\rho_i\}$  then leads to
Eq.(\ref{eq:chi0}). Again, Eq.(\ref{eq:chi1}) is obtained by using
Lagrange multipliers as above.
In fact, for unital channels (\ref{eq:chi1}) has been obtained in \cite{Stormer}.}

\section{Simplifying additivity problems}

Let us now turn to the additivity conjectures and exploit
Prop.\ref{prop:ds} in order to show that in several cases weak
additivity (for equal channels or states) implies strong
additivity (i.e., for different ones).

\begin{lemma}[Reduction for channels]\label{prop:channel1} The following (in-) 
equalities hold for arbitrary
pairs of different channels $T_1$ and $T_2$ iff they hold for
arbitrary equal pairs $T_1=T_2$.
\begin{enumerate}
\item $S_{\min,\alpha}(T_1 \otimes T_2)
=S_{\min,\alpha}(T_1)+S_{\min,\alpha}(T_2)$, for any $\alpha\geq
1$.

\item
$\chi(T_1\otimes T_2)=\chi(T_1)+\chi(T_2)$.

\item $H_{T_1 \otimes T_2}(\rho) \geq
H_{T_1}(\rho_1)+H_{T_2}(\rho_2)$ for all states $\rho$ with
respective subsystems $\rho_1, \rho_2$.

\item $H_{T_1 \otimes T_2}(\rho_1 \otimes \rho_2) =
H_{T_1}(\rho_1)+H_{T_2}(\rho_2)$ for all product states
$\rho_1\otimes\rho_2$.
\end{enumerate}
\end{lemma}

{\bf Remark:} The conjectured equality in 1. is the additivity of
the minimal output entropy when $\alpha=1$ \cite{KR01}, and it
becomes
 the multiplicativity of maximal output $p$-norms for $p=\alpha>1$. This was 
conjectured to be true for all $\alpha \in
[1,+\infty]$ before a counterexample was found \cite{WH02} ruling
out all values $\alpha>4.79$. The equation in 2. is the
conjectured additivity of the HSW capacity, which gives the
classical capacity as long as entangled states are not allowed to  
be used in the encoding \cite{Holevo,SW}. The additivity would
show that the HSW capacity itself is the unconstrained classical
capacity of quantum channels. The conjectures 3. and 4. are called
strong superadditivity and additivity of the convex closure of the
output entropy. When $T_1, T_2$ are partial traces they become
strong superadditivity and additivity of entanglement of
formation, respectively, which we discuss in greater detail below.

We note that Prop.\ref{prop:channel1} remains valid in the case
where `arbitrary channels' refers to a restricted set of channels
which is closed under direct sums and tensor products.

\proof{1. Let $\sigma_1$ and $\sigma_2$ be optimal output states
for $T_1$ and $T_2$ respectively. Then, form the following two
channels: \be T_1^\prime(\rho)= T_1(\rho) \otimes \sigma_2,\qquad
T_2^\prime(\rho)= \sigma_1 \otimes T_2(\rho). \ee It is not
difficult to see that $T_1 \otimes T_2$ and $T_1^\prime \otimes
T_2^\prime$ share the additivity property. Hence we can assume
that $T_1$ and $T_2$ have the same optimal output:
$S_{\min,\alpha}(T_1)=S_{\min,\alpha}(T_2)$.
If we apply first weak additivity and then Prop. 1.1. we obtain:
\begin{align}
&S_{\min,\alpha}(((T_1 \oplus T_2) \otimes (T_1 \oplus T_2))) \label{eq:Smin2}\\
&= 2S_{\min, \alpha}(T_1 \oplus T_2) = S_{\min,\alpha}(T_1) +
S_{\min,\alpha} (T_2).\label{eq:Smin3}
\end{align}
On the other hand, if we first apply Prop. 1.1. and then weak
additivity, we obtain that (\ref{eq:Smin2}) and thus
(\ref{eq:Smin3}) is upper bounded by $S_{\min,\alpha}(T_1\otimes
T_2)$. The converse inequality is trivial.
 \vspace*{5pt}

2. Consider
\begin{align}
&\chi \Big[\big(T_1\oplus T_2\big)^{\otimes2}\Big]
= 2\chi \big(T_1\oplus T_2\big)
=2\log\Big[2^{\chi(T_1)}+2^{\chi(T_2)}\Big]
\nonumber\\
&\qquad\qquad\qquad
=\log\Big[2^{2\chi(T_1)}+2^{2\chi(T_2)}+2^{\chi(T_1)+\chi(T_2)+1}\Big].
\nonumber
\end{align}
This follows from first
applying weak additivity and then the proposition 1.4.
On the other hand, applying them in reverse order we have
\begin{align}
&\chi \Big[\big(T_1\oplus T_2\big)^{\otimes2}\Big] \notag\\
&\qquad= \log\Big[2^{\chi(T_1 \otimes T_1)}+2^{\chi(T_2 \otimes T_2)}
+2\cdot 2^{\chi(T_1\otimes T_2)}\Big]\nonumber \\
&\qquad= \log\Big[2^{2\chi(T_1)}+2^{2\chi(T_2)}
+2^{\chi(T_1\otimes T_2)+1}\Big].\nonumber
\end{align}
Together they prove the claimed equality.
\vspace*{5pt}

For 3. we obtain by weak superadditivity,
\begin{align}
H_{T_1 \otimes T_2}(\rho)&=H_{(T_1 \oplus T_2) \otimes (T_1 \oplus T_2)}(0 
\oplus \rho \oplus 0 \oplus 0) \notag\\
&\geq H_{T_1 \oplus T_2}(\rho_1 \oplus 0) + H_{T_1 \oplus T_2}(0 \oplus 
\rho_2)\notag\\
&=H_{T_1}(\rho_1) + H_{T_2}(\rho_2).
\end{align}
Here, $\rho_1, \rho_2$ are reduced states of $\rho$. This proves
3. and the statement 4. follows in a similar way when replacing
$\rho$ by a product state. }

\begin{lemma}[Unital channels]
Proving one of the conjectures in proposition 2 for all pairs of
identical unital channels would show the conjecture is true for
arbitrary channels.
\end{lemma}

\proof{ In \cite{Fuk06} a unital channel $\tilde{T}$ is
constructed for a given channel $T$ so that these two channels
$\tilde{T}$ and $T$ share the following additivity properties: additivity
of minimal output $\alpha$-Renyi entropy, and strong
superadditivity and additivity of the convex closure of the output
entropy. Hence these conjectures can be restricted to products
$\tilde{T}_1\otimes \tilde{T}_2$ for all channels $T_1,T_2$.
As for the HSW, we have the same reduction but
for a different reason (See the remark below). 
Finally, for the above
two unital channels $\tilde{T}_1, \tilde{T}_2$ we can construct
the direct sum $\tilde{T}_1 \oplus \tilde{T}_2$ which is again a
unital channel. Then the result follows  from the proof of
proposition 2. }

{\bf Remark:}
We explain local 
relation between minimal output entropy 
and the HSW capacity, 
which was implicitly written but not clear in \cite{Fuk06}.
Since the unital extension $\tilde{T}$
sort of mixes up outputs of $T$
we have the following formula.
\be \chi( \tilde{T}_1\otimes \tilde{T}_2) = \log d_1
d_2 - S_{\rm min}( \tilde{T}_1\otimes \tilde{T}_2
)\label{eq:HSWequiv},\ee where $d_1, d_2$ are the dimensions of
the output spaces of $\tilde{T}_1$ and $\tilde{T}_2$ respectively.
Hence
the additivity of HSW capacity is 
equivalent to the additivity of the minimal output
entropy  for 
products of those extensions $\tilde{T}_1\otimes \tilde{T}_2$ 
by Eq.(\ref{eq:HSWequiv}).
Hence the additivity conjecture of the HSW capacity
can also be restricted to products $\tilde{T}_1\otimes
\tilde{T}_2$ for all channels $T_1,T_2$ by using global
equivalence \cite{Sho03,Pom03,AB04,MATWIN}. 
\vspace*{5pt}

Finally, we will discuss additivity issues for entanglement
measures. The one already mentioned is the \emph{entanglement of
formation} which was introduced in \cite{BDSW96}. Since then the
following conjectures have been considered:
\begin{align}
E_F(\rho) &\geq E_F (\rho_1) + E_F (\rho_2)\\
E_F(\rho_1 \otimes \rho_2) &= E_F (\rho_1)+ E_F
(\rho_2).\label{eq:EFadd}
\end{align}
In fact, both are again globally equivalent to the additivity of
the HSW capacity and the minimal output entropy. Moreover,
additivity would imply that $E_F$ equals an important
operationally defined entanglement measure, the \emph{entanglement
cost} $E_c$, since $E_c(\rho)=\lim_{n\rightarrow\infty}
E_F(\rho^{\otimes n})$ \cite{Ecost}.

 The entanglement
of formation $E_F(\rho)$ is the convex closure of output entropy
$H_T(\rho)$ when T is a partial trace.

Following a similar strategy as above we will now show that strong
additivity in the sense of Eq.(\ref{eq:EFadd}) is again implied by
weak additivity (i.e., Eq.(\ref{eq:EFadd}) with $\rho_1=\rho_2$).
In fact, this will not only hold for $E_F$ but for any convex
entanglement monotone \cite{BDSW96,VedralPlenio97,Vidal98}. The
main reason behind is that every such functional satisfies
\cite{Horodecki04}: \be\label{entanglement_affine} f(\oplus_i
\lambda_i \rho_i)= \sum_i \lambda_i f(\rho_i), \ee where
$\{\lambda_i\}$ is a probability distribution and $\rho_i$ are
states as before.
\begin{lemma}[Convex entanglement monotones]\cite{ERATO}
Suppose $f$ is a convex entanglement monotone which
is weakly additive, i.e., $f(\rho_1\otimes
\rho_2)=f(\rho_1)+f(\rho_2)$ for all $\rho_1=\rho_2$. Then $f$ is
strongly additive in the sense that this holds also for all
$\rho_1\neq\rho_2$.
\end{lemma}

\proof{
Let $\rho=\frac{1}{2}(\rho _1\oplus \rho_2)$. Then
\be
f(\rho \otimes \rho)
= 2f(\rho) =f(\rho_1)+ f(\rho_2).
\ee
Here, we applied the weak additivity and then (\ref{entanglement_affine}).
Applying them in reverse order we get
\begin{align}
f(\rho \otimes \rho)
&= \frac{1}{4}\left(
\sum_{i,j=1}^2 f(\rho_i \otimes \rho_j)\right)\notag\\
&= \frac{1}{2}(f(\rho_1) + f(\rho_2) + f(\rho_1 \otimes \rho_2)).
\end{align}
}

 Using similar ideas, it has recently been shown that for
regularized  entanglement measures like $E_c$ or the asymptotic
relative entropy of entanglement, monotonicity (i.e., essentially
Eq.(\ref{entanglement_affine})) and strong additivity are
equivalent \cite{Plenio}.

{\bf Acknowledgement} 
M.F. would like to thank his supervisor Y.M.Suhov for
constant encouragement and  numerous discussions. 
M.W. thanks K.G.
Vollbrecht for discussions and J.I. Cirac for  support.
Both authors thank M. B. Ruskai for bringing \cite{Stormer} to
their attention.

\end{document}